\documentclass[12pt]{article}
\usepackage{graphicx}
\usepackage{subfigure}
\usepackage{amsfonts}
\usepackage{amssymb,amsmath}

\newcommand{\bea}{\begin{eqnarray}}
\newcommand{\eea}{\end{eqnarray}}

\usepackage{multicol}
\usepackage[usenames,dvipsnames]{xcolor}
\definecolor{rosso}{cmyk}{0,1,1,0.3}
\definecolor{verde}{cmyk}{0.8,0,0.6,0.25}
\definecolor{bluc}{cmyk}{1,0.4,0,0.1}
\definecolor{blucc}{cmyk}{0.8,0.3,0,0}

\def\be{\begin{equation}}
\def\ee{\end{equation}}

\def\({\left(}
\def\){\right)}
\def\1{^{(1)}}
\def\2{^{(2)}}

\def\<{\langle}
\def\>{\rangle}

\begin{document}

\begin{titlepage}

\begin{flushright}
UT-13-05\\
IPMU 13-0029
\end{flushright}

\vskip 3cm

\begin{center}

{\large \bf 
Anomaly mediation deformed by axion
}

\vskip .5in

{
Kazunori Nakayama$^{(a,b)}$
and
Tsutomu T. Yanagida$^{(b)}$
}

\vskip .3in

{\em
$^a$Department of Physics, University of Tokyo, Bunkyo-ku, Tokyo 113-0033, Japan \vspace{0.2cm}\\
$^b$Kavli Institute for the Physics and Mathematics of the Universe,
University of Tokyo, Kashiwa 277-8583, Japan \\
}

\end{center}

\vskip .5in

\begin{abstract}

We show that in supersymmetric axion models 
the axion supermultiplet obtains a sizable $F$-term due to a non-supersymmetric dynamics
and it generally gives the gaugino masses comparable to the anomaly mediation contribution.
Thus the gaugino mass relation predicted by the anomaly mediation effect can be significantly modified
in the presence of axion to solve the strong CP problem.

\end{abstract}

\end{titlepage}

\setcounter{page}{1}


Pure gravity mediation models~\cite{Ibe:2011aa,arkani} were proposed just after the discovery of the Higgs boson of mass $\sim 125$ GeV at the LHC~\cite{:2012gk}. The gravitino mass is assumed to be $100 - 1000$\,TeV, and squarks and sleptons are also as heavy as the gravitino. 
The relatively large mass of the Higgs boson is easily explained by radiative corrections of top-quark loops as pointed out in \cite{Okada:1990vk,Giudice:2011cg}. The gauginos obtain their masses of around $100 - 1000$\,GeV 
through the anomaly mediated SUSY breaking (AMSB) effect~\cite{Giudice:1998xp,Randall:1998uk,Dine:1992yw}. 
Since we do not introduce the Polonyi field to give gaugino masses, there is no cosmological Polonyi problem~\cite{Coughlan:1983ci,Nakayama:2011wqa}. 
The large mass of gravitino ameliorates substantially many phenomenological problems in the SUSY standard model~\cite{Ibe:2011aa}.

To test the pure gravity mediation at the LHC, the gaugino mass spectrum is very crucial as shown in \cite{Ibe:2011aa,Moroi:2011ab}. 
In this letter, we show that the gaugino mass relation given by the AMSB effect is naturally deformed in simple axion models.
We will see that this is rather a generic feature in SUSY axion models, since the axion multiplet necessarily couples to
the gauge field strength in order to solve the strong CP problem 
and the axion multiplet generally obtains a sizable $F$-term due to the SUSY braking effect.
This acts as a gauge-mediation effect, which substantially modifies the gaugino mass spectrum.


First, let us consider a SUSY axion model~\cite{Kawasaki:2013ae}, where the superpotential is given by
\begin{equation}
	W = \lambda X(\Phi_1\Phi_2-v^2) + k \Phi_1 Q\bar Q + W_0.
	\label{super}
\end{equation}
Here $X, \Phi_1, \Phi_2$ are gauge singlet chiral superfield and they have charges $0, +1, -1$ 
under the global Peccei-Quinn (PQ) symmetry U(1)$_{\rm PQ}$,
and R charges $+2, 0, 0$, respectively. 
The PQ quarks $Q$ and $\bar Q$, which are fundamental and anti-fundamental representations of color SU(3) respectively,
both have PQ charges $-1/2$.
Since the global U(1)$_{\rm PQ}$ symmetry is anomalous under the QCD, the axion appears and the strong CP problem
is solved~\cite{Peccei:1977hh}.
Constants $\lambda$, $k$, $v$ and $W_0$ are taken to be real and positive without loss of generality,
and $W_0=m_{3/2}M_P^2$ for vanishing cosmological constant with $m_{3/2}$ being the gravitino mass.
The K\"ahler potential is given by 
\begin{equation}
	K =|\Phi_1|^2+|\Phi_2|^2+|X|^2 - c_1'\frac{|\Phi_1|^2|z|^2}{M_P^2} - c_2'\frac{|\Phi_2|^2|z|^2}{M_P^2} + \dots,
	\label{kahler}
\end{equation}
where $z$ denotes the SUSY breaking field, $M_P$ the reduced Planck scale and $c_1'$ and $c_2'$ are real constants of order unity.
Note that $z$ should be charged under some symmetry to forbid a direct coupling to gauginos,
as is usual in dynamical SUSY breaking models,
and hence terms linear in $z$ do not appear in (\ref{kahler}).
Dots represent higher order terms that are irrelevant for our discussion.
Including the SUSY breaking effect, the scalar potential reads
\begin{equation}
\begin{split}
	V =& \lambda^2 |\Phi_1\Phi_2-v^2|^2 + \lambda^2|X|^2(|\Phi_1|^2+|\Phi_2|^2) + 2\lambda m_{3/2}v^2(X+X^\dagger) \\
	    &+ c_1m_\phi^2|\Phi_1|^2 + c_2 m_\phi^2|\Phi_2|^2,
\end{split}
\end{equation}
where $c_1m_\phi^2 \equiv (3c_1'+1)m_{3/2}^2$ and $c_2 m_\phi^2 \equiv (3c_2'+1)m_{3/2}^2$, both are assumed to be positive.
By minimizing the scalar potential, we find\footnote{
	While overall phase of the combination $\Phi_1\Phi_2$ is fixed to be zero by minimizing the potential, 
	the relative phase between $\Phi_1$ and $\Phi_2$, corresponding to the massless axion, is not fixed.
	For simplicity, in the following we take a basis such that $\langle\Phi_1\rangle$ is real and positive.
	Otherwise, a phase factor appears in (\ref{eps}). However, it does not affect the main results.
}
\begin{equation}
\begin{split}
	&v_X\equiv \langle X\rangle = -\frac{2m_{3/2} v^2}{\lambda(v_1^2+v_2^2)}, \\
	&v_1 \equiv \langle |\Phi_1|\rangle = v\left( \frac{c_2 m_\phi^2+\lambda^2 v_X^2}{c_1 m_\phi^2+\lambda^2 v_X^2} \right)^{1/4}
		+ \mathcal O\left(\frac{c_1m_\phi^2}{v}\right),\\
	&v_2 \equiv \langle |\Phi_2|\rangle = v\left( \frac{c_1 m_\phi^2+\lambda^2 v_X^2}{c_2 m_\phi^2+\lambda^2 v_X^2} \right)^{1/4}
		+ \mathcal O\left(\frac{c_2m_\phi^2}{v}\right).
\end{split}
\end{equation}
At this minimum, the PQ scalar $\Phi_1$ obtains a $F$-term as
\begin{equation}
	F^{\Phi_1} = -e^{K/M_P^2}K^{\Phi_1\Phi_1^*}\left(W_{\Phi_1}+ \frac{K_{\Phi_1}W}{M_P^2}\right)^\dagger =  -m_{3/2}v_1 \epsilon,
	\label{Fphi}
\end{equation}
where $K^{\Phi_i\Phi_j^*} = (K_{\Phi_i\Phi_j^*})^{-1}$ and subscript $\Phi_i$ 
denotes the derivative with respect to $\Phi_i$, and\footnote{
	$F^{\Phi_1}$ also receives contribution from $K^{\Phi_1 z^*}(D_z W)^\dagger$,
	which is same order if $\langle z \rangle \sim M_P$. In dynamical SUSY breaking scenario, however,
	$\langle z \rangle$ is sufficiently small and such a contribution can be neglected.
}
\begin{equation}
	\epsilon = \frac{c_2-c_1}{c_2+c_1+2\lambda^2 v_X^2/m_\phi^2}.
	\label{eps}
\end{equation}
Note that the $F$-term at this order $(\sim \mathcal O(m_{3/2}v))$ vanishes for $c_1=c_2$ in which case $v_1=v_2\simeq v$.
In general, however, $c_1$ and $c_2$ are free parameters of order unity, hence such a cancellation does not occur
and $\epsilon$ is generically $\mathcal O(1)$ (positive or negative).\footnote{
	At low energy, the mass term for $\Phi_1$, $c_1 m_\phi^2$, receives negative contribution from the coupling
	$k\Phi_1 Q\bar Q$ through the renormalization group evolution.
	Therefore, even if we assume $c_1=c_2$ at high energy, the relation $c_1 < c_2$ holds at low energy.
	In particular, if $k\sim \mathcal O(1)$, we generally obtain $c_2 - c_1\sim \mathcal O(1)$, which then 
	predicts positive $\epsilon \sim \mathcal O(1)$.
}
This is understood as a consequence of SUSY breaking effect :
without SUSY breaking effect with non-zero constant term $W_0$, 
$v_1=v_2\simeq v$ corresponds to the SUSY minimum although the vacuum energy is negative there.
After turing on the SUSY breaking effect, the SUSY vacuum is lifted up but the position of the minimum changes
with an amount of $\mathcal O(v)$ along the flat direction, where the PQ scalars obtain $F$-terms.

Below the PQ scale, after integrating out heavy PQ quarks, we obtain the following effective Lagrangian~\cite{Kim:1979if},
\begin{equation}
	-\mathcal L = \sum_{i=1}^3\frac{\mathcal C_i\alpha_i}{8\pi} \int d^2\theta \ln(\Phi_1) \mathcal W_i^a\mathcal W_i^a +{\rm h.c.},
	\label{gaugino}
\end{equation}
where $\alpha_i$ are the fine-structure constants of SM gauge groups with $i=1,2,3$ corresponding to U(1), SU(2) and SU(3),
$\mathcal W_i^a$ are SUSY gauge field strength,
and $\mathcal C_i$ are constants which depend on gauge charge assignments on PQ quarks.
For example, if we introduce $N_5$ pairs of PQ quarks with 5 and $\bar 5$ representations of SU(5), 
we have $\mathcal C_1=\mathcal C_2=\mathcal C_3 = N_5$.
Considering the mixing of $\Phi_1$ and $\Phi_2$ into mass eigenstates around the vacuum , 
we find that the PQ scale, often denoted by $f_a$,
is given by $f_a = \sqrt{2(v_1^2+v_2^2)} / \mathcal C_3$ with $\mathcal C_3$ corresponding to the domain wall number.
The term (\ref{gaugino}) generates the gaugino masses, $\delta M_\lambda^i$, due to the non-zero $F$-term of $\Phi_1$ as
\begin{equation}
	\delta M_\lambda^i (M_{\rm PQ}) = -\mathcal C_i\frac{\alpha_i}{4\pi}\frac{F^{\Phi_1}}{v_1} = \mathcal C_i\frac{\alpha_i}{4\pi} m_{3/2} \epsilon.
	\label{Gmass}
\end{equation}
at the PQ messenger scale, $M_{\rm PQ} \equiv k v_1$.
It is notable that the gaugino masses induced by the PQ scalar (\ref{Gmass}) is comparable to the AMSB contribution.
Combining them with the AMSB contribution, we obtain the gaugino masses as (see Appendix)
\begin{equation}
	M_\lambda^i(M_{\rm PQ}) = \frac{\alpha_i}{4\pi}(-b_i + \mathcal C_i\epsilon) m_{3/2}
	= \begin{cases}
		\frac{\alpha_1}{4\pi}\left(\frac{33}{5} + \mathcal C_1\epsilon \right) m_{3/2} &~{\rm for~bino}\\
		\frac{\alpha_2}{4\pi}\left(1 + \mathcal C_2\epsilon \right) m_{3/2} &~{\rm for~wino}\\
		\frac{\alpha_3}{4\pi}\left(-3 + \mathcal C_3\epsilon \right) m_{3/2} &~{\rm for~gluino}.
	\end{cases}
	\label{M_lam}
\end{equation}
Therefore, for positive $\epsilon$, the gluino tends to be light while the wino and bino tend to become massive.
In the following, as a concrete example, we take $\mathcal C_1=\mathcal C_2=\mathcal C_3 = N_5$ by introducing 
$N_5$ pairs of PQ quarks with 5 and $\bar 5$ representations of SU(5).\footnote{
	For solving the strong CP problem, only non-zero $\mathcal C_3$ is needed.
	However, introducing $5$ and $\bar 5$ pair is favored in order to maintain the successful gauge coupling unification.
}

\begin{figure}
\begin{center}
\vskip -1.cm
\includegraphics[scale=1.2]{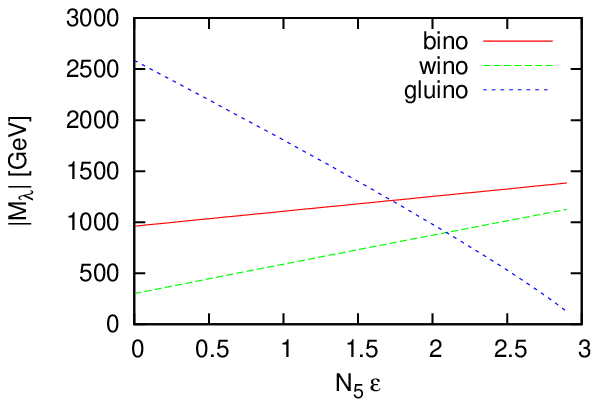}
\includegraphics[scale=1.2]{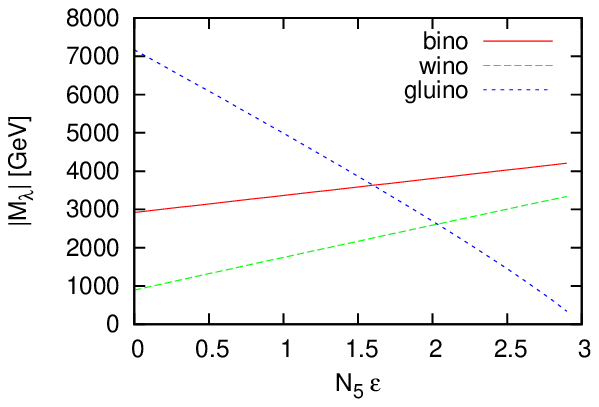}
\caption{Gaugino masses as a function of $N_5 \epsilon$ for $m_{3/2}=100$\,TeV (left) and $m_{3/2}=300$\,TeV (right).
We have taken all scalar masses, as well as the higgsino masses, to be equal to $m_{3/2}$.}
\label{fig:mass}
\end{center}
\end{figure}

Fig.~\ref{fig:mass} shows gaugino masses as a function of $N_5 \epsilon$ for $m_{3/2}=100$\,TeV (left) 
and $m_{3/2}=300$\,TeV (right) at the one-loop level.
We have taken all scalar masses, as well as the higgsino masses, to be equal to $m_{3/2}$.
The result is insensitive to the choice of $M_{\rm PQ}$.
It is clearly seen that gaugino mass relations are significantly modified for $N_5 \epsilon \sim \mathcal O(1)$.
Note that we have ignored the higgsino mass threshold correction, which could further modify the mass spectrum.
Theoretically we naturally expect $N_5 \epsilon \sim \mathcal O(1)$.
It may be welcome since the gluino mass is much lighter than the prediction of pure-gravity mediation,
which may make the detection of gluino at the LHC easier.

Next, we consider another SUSY axion model~\cite{Murayama:1992dj,Nakayama:2012zc}. The superpotential is given by
\begin{equation}
	W = \frac{\Phi_1^n \Phi_2}{M^{n-2}} + k\Phi_1 Q\bar Q + W_0, 
\end{equation}
where $\Phi_1, \Phi_2$ are gauge singlet chiral superfield and they have charges $+1, -n$ under U(1)$_{\rm PQ}$,
$M$ is the cutoff scale, $n\geq 3$ is an integer, and $Q$ and $\bar Q$ are PQ quarks as in the previous model.
Taking account of the SUSY breaking effect, the scalar potential is given by
\begin{equation}
	V = \frac{|\Phi_1|^{2n-2}}{M^{2n-4}}(|\Phi_1|^2+n^2 |\Phi_2|^2) +(n-2)m_{3/2}\left( \frac{\Phi_1^n \Phi_2}{M^{n-2}}+{\rm h.c.} \right)
		+ c_1 m_\phi^2 |\Phi_1|^2 + c_2 m_\phi^2 |\Phi_1|^2.
\end{equation}
We assume $c_1 m_\phi^2 <0$. The minimum of the potential is found to be\footnote{
	Again, although the overall phase of $\Phi_1^n \Phi_2$ is fixed by minimizing the potential,
	the relative phase between $\Phi_1$ and $\Phi_2$, corresponding to the axion, remains undetermined.
	Here we choose real and positive $\langle\Phi_1\rangle$ without loss of generality.
}
\begin{equation}
\begin{split}
	&v_1^{2n-2} \equiv \langle |\Phi_1|\rangle^{2n-2} \simeq \left(-c_1m_\phi^2 + \frac{(n-2)^2}{n^2}m_{3/2}^2 \right)\frac{M^{2n-4}}{n},\\
	&v_2 \equiv \langle |\Phi_2|\rangle \simeq v_1 \frac{(n-2)m_{3/2}}{n^{3/2}\left( -c_1m_\phi^2+\frac{(n-2)^2}{n^2}m_{3/2}^2 \right)^{1/2}}.
\end{split}
\end{equation}
Here we have assumed $|c_2 m_\phi^2|\ll |c_1m_\phi^2|$ for simplicity.
For $n=3$, for example, $v_1 \sim \sqrt{m_\phi M}$ and it is $\mathcal O(10^{12}\,{\rm GeV})$ for $m_\phi \sim 100$\,TeV
and $M\sim M_P$, hence desirable PQ scale is obtained.
At this minimum, $\Phi_1$ obtains a $F$-term as
\begin{equation}
	F^{\Phi_1} = -\frac{2}{n}m_{3/2}v_1.
\end{equation}
Therefore, it affects the gaugino masses as in (\ref{M_lam}) with $\epsilon = 2/n$.
These examples show that modification on the gaugino masses in the SUSY axion model is a generic feature,
if the gaugino masses are mainly given by AMSB contribution.

Let us briefly discuss cosmological aspects of the PQ models.
In both models, the axino mass ($m_{\tilde a}$) is comparable to the gravitino mass and hence it is much heavier than gauginos.
The axino decay rate into gluino and gluon is given by~\cite{Choi:2008zq}
\begin{equation}
	\Gamma(\tilde a \to \tilde g+g)=\frac{\alpha_3^2}{16\pi^3}\frac{m_{\tilde a}^3}{f_a^2}\left(1-\frac{m_{\tilde g}^2}{m_{\tilde a}^2} \right)^3.
\end{equation}
Thus the decay temperature of the axino is estimated as
\begin{equation}
	T_{\tilde a} \sim 4\times 10^3\,{\rm GeV}\left( \frac{m_{\tilde a}}{10^3\,{\rm TeV}} \right)^{3/2}
	\left( \frac{10^{12}\,{\rm GeV}}{f_a} \right).
\end{equation}
If this is larger than the wino decoupling temperature, the axino does not cause any cosmological problem.
On the other hand, if this is slightly smaller than the decoupling temperature, winos produced by the axino decay
annihilate and may result in correct dark matter abundance.
If $T_{\tilde a} \lesssim 1$\,GeV, winos are likely to be overproduced or they must be too light to conflict with 
astrophysical/cosmological constraints~\cite{Hisano:2008ti}.
The axino abundance, in terms of the energy density-to-entropy density ratio, is given by~\cite{Brandenburg:2004du}
\begin{equation}
	\frac{\rho_{\tilde a}}{s} \sim 9\times 10^{-1}\,{\rm GeV}g_3^6\ln\left(\frac{3}{g_3}\right)
	\left( \frac{m_{\tilde a}}{10^3\,{\rm TeV}} \right)
	\left( \frac{T_{\rm R}}{10^7\,{\rm GeV}} \right)
	\left( \frac{10^{12}\,{\rm GeV}}{f_a} \right)^2,
\end{equation}
where $T_{\rm R}$ denotes the reheating temperature after inflation.
Therefore, axinos decay before they come to dominate the Universe for reasonable parameters : e.g., $T_{\rm R}\sim 10^9$\,GeV
and $f_a \sim 10^{12}$\,GeV. Hence thermal leptogenesis works successfully~\cite{Fukugita:1986hr}.
Saxions are also produced thermally~\cite{Graf:2012hb}, whose cosmological aspects are similar to the axino discussed above.
For such parameters, the saxion coherent oscillation abundance is smaller than the thermal one~\cite{Kawasaki:2007mk}.\footnote{
	Thermal effects on the saxion coherent oscillation dynamics can be neglected 
	if $\alpha_3 T_{\rm max}^2/f_a \lesssim m_\phi$~\cite{Kawasaki:2010gv,Moroi:2012vu}.
	Here $T_{\rm max}$ is the maximum temperature after inflation : $T_{\rm max} \sim (T_{\rm R}^2 H_{\rm inf} M_P)^{1/4}$,
	with $H_{\rm inf}$ being the Hubble scale during inflation. This can be satisfied for low-scale inflation.
	Actually we need low-scale inflation in order to avoid too large axion isocurvature fluctuation~\cite{Kawasaki:2008sn}.
}


Let us mention relation to other works.
A similar effect on the sparticle masses from the axion multiplet was discussed in Ref.~\cite{Abe:2001cg}
in the context of AMSB in which observable sector is sequestered from the SUSY breaking sector.
There it was mentioned that the model (\ref{super}) does not affect sparticle masses.
This is because their situation corresponds to $c_1=c_2=0$ due to the sequestering.
In the pure gravity-mediation, there is no reason for that.
Ref.~\cite{Choi:2011xt} considered such effects in the models with anomalous U(1) gauge symmetry.
Recently, Ref.~\cite{Baryakhtar:2013wy} discussed so-called axion mediation motivated by the string theory.
However, the axion $F$-term was taken as a free parameter.
In the so-called deflected AMSB model~\cite{Pomarol:1999ie}, a messenger sector was introduced
to modify the AMSB mass spectrum in a similar way~\cite{Okada:2012nr}.
While it is a general setup, explicit constructions based on the PQ model have been missing.

Finally we emphasize that the coupling like (\ref{gaugino}) must exist if the strong CP problem is solved by the axion
independently of the details of the PQ model.
In particular, in string theory, K\"ahler moduli super multiplets can have such couplings to the visible sector gauge fields
if we live on D-brane spanning the three-dimensional space in the type IIB theory.
Thus one of such moduli may take a role of PQ scalar $\Phi$ whose imaginary component,
originating from the 10-dimensional Ramond-Ramond $p$-form field, corresponds to the QCD axion~\cite{Conlon:2006tq,Conlon:2006ur,Choi:2006za,Acharya:2010zx,Cicoli:2012sz}.
Here it should be noticed that the real part of $\Phi$, the saxion, must be stabilized in a non-supersymmetric way
since otherwise the saxion stabilization would give the axion mass similar to the saxion, which spoils the PQ solution to the strong CP problem.
Then it seems to be plausible that $\Phi$ obtains a $F$-term of order $\sim m_{3/2} f_a$ and hence it gives
corrections to the gaugino masses which are same order of the AMSB contribution
(see also \cite{Conlon:2006ur,Choi:2006za,Higaki:2011me}).\footnote{
	If the saxion is stabilized at the Planck scale, it may have an $F$-term of $\sim m_{3/2}M_P$,
	and actually it acts as the Polonyi-like field which dominates over the AMSB contribution.
}
Therefore, if we believe the axion as a solution to the strong CP problem, we should be careful not only on the saxion stabilization
but also on its effects on gaugino masses.

\section*{Acknowledgment}

T.T.Y. thanks J. Evans, M. Ibe and K. A. Olive for a useful discussion on pure gravity mediation models.
This work was supported by JSPS KAKENHI Grant 
No.\ 22244030 (K.N.) and by the MEXT Grant-in-Aid No.\ 21111006 (K.N.).
This work was supported by World Premier International Research Center Initiative (WPI Initiative), MEXT, Japan.

\section*{Appendix}

Here we derive Eq.~(\ref{M_lam}).
In the superconformal language~\cite{Cremmer:1978hn}, the total Lagrangian involving the chiral superfield $\Phi_i$ is given by
(in this Appendix we take the unit $M_P=1$)
\begin{equation}
	\mathcal L = \int d^4\theta C^\dagger C \Omega(\hat \Phi_i,\hat \Phi_i^\dagger) 
	+ \left[\int d^2\theta \left(C^3 W (\hat \Phi_i) + \frac{1}{4}f(\hat \Phi_i) \mathcal W^a \mathcal W^a \right)+{\rm h.c.}\right],
\end{equation}
where $C$ is a compensator field, $\hat \Phi_i \equiv \Phi_i/C$ and $\Omega = -3e^{-K/3}$.
Taking $C = e^{K/6}$, we find $F^C/C = F^{i} K_{i}/3 + m_{3/2}$
where $F^{i}=-e^{K/2}K^{i\bar j} D_{\bar j}\bar W$, and it recovers the Einstein frame action.
Here the subscript $i$ should be regarded as that for $\hat \Phi_i$.
The gaugino mass is read from the fact that the lowest component of the gauge kinetic function corresponds to the running
gauge coupling : $f(\mu, \hat \Phi_i) = 1/g^2(\mu, \hat \Phi_i)$ at the scale $\mu$.
The anomaly-mediation effect can be incorporated via the replacement $\mu \to \hat\mu \equiv \mu / \sqrt{C^\dagger C}$~\cite{Pomarol:1999ie}.
Let us assume that there is only one chiral superfiled, $\Phi$, which enters the running of $g$.
As in the case of axion model, $\Phi$ couples to $N_5$ pairs of vector-like chiral superfields through $W = k\Phi Q\bar Q$
where $Q$ and $\bar Q$ are assumed to be $5$ and $\bar 5$ of SU(5).
Therefore, the running of the SU$(N_c)$ gauge coupling reads
\begin{equation}
	\frac{dg}{d\ln\mu} = \begin{cases} 
		-\frac{g^3}{16\pi^2} b &~{\rm for}~\mu < k\Phi,\\
		-\frac{g^3}{16\pi^2} b' &~{\rm for}~\mu > k\Phi,\\
	\end{cases}
\end{equation}
where $b=3N_c-N_f$ and $b'=3N_c-N_f-N_5$ with $N_f$ counting number of (anti-)fundamental representations of SU$(N_c)$ 
with weight $1/2$. Now the gaugino mass at the scale $\mu < k\Phi$ is given by
\begin{equation}
\begin{split}
	M_\lambda (\mu) &= F^C\partial_C \ln ({\rm Re} f(\hat\mu, \hat \Phi) ) +   F^{\hat\Phi}\partial_{\hat\Phi} \ln ({\rm Re} f(\hat\mu, \hat \Phi) )\\
	&= -\frac{g^2(\mu)}{16\pi^2}\left(b \frac{F^C}{C} +(b-b')\frac{F^{\hat\Phi}}{\hat\Phi}\right) \\
	&=  -\frac{\alpha(\mu)}{4\pi}\left(b m_{3/2}+N_5\frac{F^{\hat\Phi}}{\hat\Phi}\right).
	\label{M_lam_mu}
\end{split}
\end{equation}
The first term corresponds to the AMSB contribution and the second term corresponds to the threshold correction
from the decoupling of $Q$ and $\bar Q$, similar to the gauge-mediation.
This reproduces Eq.~(\ref{M_lam}).
By assuming that the gaugino mass at the scale $\Lambda > k\Phi$ is given by the purely AMSB form, 
we can rewrite (\ref{M_lam_mu}) as
\begin{equation}
	\frac{M_\lambda(\mu)}{g^2(\mu)} = \frac{M_\lambda(\Lambda)}{g^2(\Lambda)} 
	- \frac{N_5}{16\pi^2}\left(m_{3/2}+\frac{F^{\hat\Phi}}{\hat\Phi}\right).
\end{equation}
If $F^{\hat\Phi}=0$, we recover the AMSB result at low energy : $M_\lambda(\mu)/g^2(\mu)=-(b/16\pi^2)m_{3/2}$.
This corresponds to the UV insensitive property of the AMSB~\cite{Giudice:1998xp,Pomarol:1999ie}.
If $F^{\hat\Phi}\neq 0$, on the other hand, the low-energy prediction on the gaugino mass is modified from the purely AMSB one.



\end{document}